# Teardown and feasibility study of IronKey – the most secure USB Flash drive

Sergei Skorobogatov
*Computer Science and Technology*
*University of Cambridge*
Cambridge, United Kingdom
sps32@cam.ac.uk

*Abstract*—There are many solutions for protecting user data on USB Flash drives. However, the family of IronKey devices was designed with the highest security expectations. They are definitely standing above others by being certified to FIPS 140-2 Level 3 and also claimed as certified by NATO for Top-Secret use. Many encrypted USB drives had been evaluated and found insecure, however, no public research on IronKey devices was made. This feasibility study fills the gap by looking inside the IronKey family of devices. As a result the users of the IronKey devices could be assured about the real level of the security protection they get. Several generations of devices from IronKey family and competitors are teared down, their hardware solutions discussed and evaluated for possible attacks. Some potential flaws are exposed and those findings are likely to stimulate further research into specific solutions aimed to protect user data.

I. INTRODUCTION

Local storage solutions such as USB Flash drives are absolutely essential for corporate, government and military use. This is because users either cannot trust cloud storage or the devices which provide online access could be compromised. Hence, data confidentiality, integrity and authenticity of Flash drives are paramount. Encrypted USB Flash drives use hardware-based encryption and often meet a variety of worldwide security regulations. Some devices have additional hardware to support user authentication and limit the number of failed consecutive password attempts. In addition these devices address other security issues associated with standard USB Flash drives such as data leakage and computer viruses, malware and spyware.

There exist dozens of encrypted USB Flash drives models. They vary in sizes, capacity and protection level. However, they have one thing in common – the data stored inside the Flash memory is always encrypted and the access is protected with a user password. There are some challenges associated with encrypted USB storage devices. First is the need for on-the-fly encryption and decryption which cause inevitable latency and slower data transfer rate. Second is the additional cost for specialised hardware that not only performs the cryptographic operations, but also keeps the encryption key secret. Finally, there should be solutions for limiting the number of consecutive incorrect password attempts. While the standard USB Flash drive products started to emerge from early 2000, several years have passed until secure USB drives hit the market. IronKey Inc. was among the first who decided to develop such products primarily aiming at military, government and corporate users [1].

Throughout my research I have been looking at many challenging devices with very strong security claims. As a result some of those claims were smashed. This, for example, happened to military grade FPGAs [2][3], FBI claims on iPhone 5c [4], security claims about Flash storage [5], and de-processing of live ICs [6]. In that respect the IronKey Inc. claim that their devices are "World's Most Secure Flash Drive" [7] also sounded quite strong and definitely needed an independent security evaluation. In case of any findings this could be a good candidate for our popular hardware security practical, since the high security device we used in the past was completely broken with many details published [8]. There were very limited publications aimed at comparison of hardware security in USB Flash drives. One was carried out as a Master's thesis [9] and primarily outlines possible attack vectors, while more comprehensive one was done by researchers at Google [10] who demonstrated several weaknesses of some devices. However, none of those publications involved IronKey devices. Of course, many IronKey drives were certified to the high FIPS140-2 Level 3 standard and even approved by NATO for Top-Secret use according to this source [11]. Nevertheless, it would be beneficial to compare their hardware with encrypted USB Flash drives from other manufacturers. Not only this would give an assurance to the existing users, but also could help in avoiding possible data leakages in the future. The user password protection of IronKey device had been recently in the news when a legitimate user was unable to unlock a huge value of cryptocurrency assets due to forgotten password [12].

There are two independent certification procedures that are normally applied to secure hardware. One is FIPS (Federal Information Processing Standards) which is a set of US Government requirements for data and its encryption [13]. Another is Common Criteria (CC) which is an international set of specifications and guidelines designed to evaluate information security products and systems [14]. While FIPS is focused on cryptography and physical tamper resistance, Common Criteria is focused on the security functions of an IT product and the correctness of specific security features. With respect to the encrypted USB Flash drives two specific FIPS standards are used to make sure that the user data are adequately protected: FIPS 197 [15] and FIPS140-2 [16]. FIPS 197 certification looks at the hardware encryption algorithms used to protect the data. FIPS 140-2 is the more advanced level of certification that includes a rigorous analysis of the physical properties. Hence, with FIPS 140-2 certified USB Flash drives both the tamper-proof design of printed circuit board (PCB) and data encryption have been approved. If a manufacturer is aiming to sell encrypted USB Flash drives to government organisations it should aim at certifying them to FIPS 140-2 Level 3. That requires the hardware to feature tamper resistance and identity-based authentication. Also private keys can only enter or leave the module in an encrypted form. The FIPS 140-2 standard is currently being replaced with the new FIPS 140-3 standard [17]. With respect to the Common Criteria its Evaluation Assurance Levels (EAL) correspond to different levels of the security evaluation. This is normally applied to semiconductor devices that perform authentication, key storage and encryption. Hence, for adequately protected USB Flash drive the semiconductor devices that store the encryption key, check user passwords and limit the number of consecutive incorrect password attempts should ideally comply with EAL4 or higher level of assurance. That means the device was methodically designed, tested and reviewed, or (semi)-formally designed, tested and verified.

This paper is organised as follows. Section 2 gives brief introduction into the history of IronKey and some similar devices. Section 3 presents challenges for accessing internal electronics of the USB drives and the results obtained for IronKey and other devices. Section 4 sets out the results at PCB and IC level teardown. Section 5 discusses possible attack methods and presents successful attacks on limited password retry attempts. Section 6 compares electronics and security of the devices, while Section 7 outlines further work. Finally the impact of the research is discussed in Section 8.

II. BACKGROUND

The IronKey technology has evolved over time. Several models with different hardware implementations and security features were developed. All devices present themselves as two drives – CD-ROM that stores documentation and user application for device management, and normal USB drive that stores user files. Until the correct password is entered the user drive is inaccessible. The number of consecutive incorrect password attempts is usually limited to ten after which the content of the user drive is irreversibly deleted.

*A. IronKey history*

The IronKey Inc. was an internet security and privacy company established in 1996. Between 2005 and 2007 it had developed a secure tamper-resistant USB Flash drive marketed as IronKey. To assist with the development of the most secure USB Flash drive they attracted US$1.4 million US government grant and raised US$6 million from investors [1]. The first IronKey device named D2 was released in 2007 followed by further revisions in 2008. These devices stored user data only in encrypted form. The AES256 encryption key was securely stored inside a dedicated highly secure microcontroller. In fact, it was encrypted with a user password and destroyed after ten consecutive incorrect password attempts. When plugged into a USB port of a computer the IronKey presented itself as a CD-ROM drive with specialised software for secure communication with the IronKey device. Upon entering the correct password access to user files was granted. IronKey drives had rugged metal casing filled with epoxy compound making them not only tamper-resistant, but also waterproof. Three new devices S100, D200 and S200 were released in 2009. They looked very similar with only some features improved and optimised hardware for faster Read/Write speed. Unlike the original IronKey devices certified to FIPS 140-2 Level 2 these were certified to Level 3. This was particularly important for selling them to government customers.

In September 2011 another storage device manufacturer Imation Corp. has reached a deal to buy the hardware business of IronKey [18].

*B. Imation history*

Storage device manufacturer Imation Corp. started selling secure USB Flash drives from 2011. Prior to buying the IronKey business Imation has acquired MXI Security from Memory Experts International in June 2011 [19]. This allowed them to add two new secure USB Flash drives Defender F100 and Defender F150 to their products. Upon completion of the IronKey purchase those drives started to be sold as IronKey F100 and F150 devices, both certified to FIPS 140-2 Level 3. Later the family was expanded with IronKey F200 device with biometric fingerprint identification.

In 2013 two new members were added to the IronKey family – S250 and D250 with improved characteristics, both with FIPS 140-2 Level 3 certification. Soon after this in 2014 Imation added a budgetary version to the family – IronKey D80 – affordable but not certified. However, all their IronKey devices had USB2.0 interface. That was their major disadvantage when compared with unencrypted USB3.0 Flash drives. Therefore, in 2015 the new IronKey S1000 was brought to the market. It had exceptional characteristics, high level of security and certified to FIPS 140-2 Level 3. In fact, it was the world's fastest encrypted USB Flash drive [20].

In February 2016 the IronKey business was split between DataLocker and Kingston. DataLocker Inc., a leading provider of encrypted solutions, acquired the IronKey Enterprise Management Services assets [21], while Kingston Digital Inc. has acquired the technology and assets of IronKey [22].

*C. Kingston history*

Manufacturer and distributor of memory products Kingston Digital Inc. is part of Kingston Technology Corp. founded in 1987. It started offering its own line of encrypted USB Flash drives in 2010. The first devices were called DataTraveler Locker and DataTraveler Vault Privacy. However, they did not have any certification. The first device that achieved FIPS 140-2 Level 2 certification was DataTraveler 4000. It hit the market in 2011. Since then more advanced versions of DataTraveler Locker and Vault Privacy devices were developed. In 2015 the first secure Flash drives with faster USB3.0/USB3.1 interface were introduced: DataTraveler 4000G2, Locker+G3 and Vault Privacy 3.0. However, only DT4000G2 was certified to FIPS 140-2 Level 3, while DataTraveler Vault Privacy 3.0 has achieved FIPS 197 certification.

Soon after acquiring the IronKey business a new member was added to the family of IronKey devices – D300. Not only it was certified to the high FIPS 140-2 Level 3 standard, but also achieved NATO certification [11]. One interesting observation could be made by comparing the FIPS 140-2 certification documents from DataTraveler DT4000G2 [23] and IronKey D300 [24]. Not only the wordings look exactly the same apart from the product names, but the pictures of the certified devices on page 9 are identical. The same refers to the firmware and software versions described in chapter 3.2.1.

In 2018 the latest model of IronKey device was introduced – D300S. According to its press release it had improved security and virtual keyboard [25].

*D. USB drives from other manufacturers*

DataLocker has its own range of encrypted USB Flash drives. The most known among them is the Sentry family. They include the old USB2.0 Sentry device certified to FIPS 140-2 Level 2, newer USB3.0 Sentry 3.0 device with FIPS 197 certification, and the latest Sentry One device certified to FIPS 140-2 Level 3.

Integral has several families of encrypted USB drives: Secure, Courier, Crypto, and Envoy. Most of them are either FIPS 197 or FIPS 140-2 certified.

MXI Security has developed its range of Stealth USB Flash drives certified to FIPS 140-2 Level 3. They rely on their custom developed hardware for the encryption and key management. Some of their devices were sold under IronKey name after acquisition by Imation.

SafeXS is another manufacturer of encrypted USB Flash drives. The family of these devices is called Protector.

iStorage uses different user interface for unlocking the USB drive. Their Datashur family of devices have a numerical keypad with the user password length varying between 7 and 15 digits. Although they all have a battery inside the case it is only used for unlocking the device prior plugging it into a USB slot. The user password is not erased upon discharging or temporarily removing the battery.

Further manufacturers include but not limited to Verbatim, Corsair, Apricorn, Kanguru, Freecom and Elecom.

### III. PCB LEVEL TEARDOWN

Many encrypted USB Flash drives have tamper resistant cases to prevent physical access to internal electronic components and to protect the devices from harsh environment and mechanical forces. This is especially important for military and government applications. Hence, most devices certified to FIPS 140-2 Level 3 have their electronic components encapsulated in epoxy compound.

By taking various devices apart, or performing their teardown, not only their internal hardware structure could be learned. The hardware solutions and changes in their hardware might explain certain successes and failures throughout their life.

*A. Opening up the case*

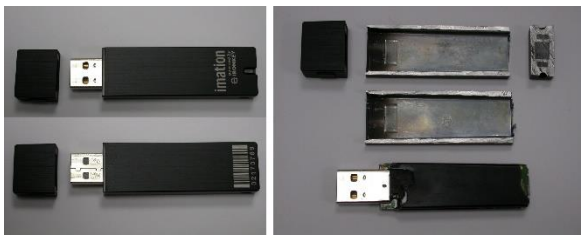

Fig. 1.  IronKey device before and after opening its metal case

The first step in accessing PCB of electronic devices is opening their enclosure. If it is made from plastic then this could be relatively easy. However, some metal cases could require cutting tools to open them up. In case of IronKey devices metal cases were cut using low-cost engraving tools with stone cutting disc. Once groves along each side of the case are created, the shells of the case could be removed. Some cases could be easier to open by cutting off the end and pushing out the PCB. Figure 1 shows the IronKey before and after opening the metal case using mechanical tools.

*B. Removal of epoxy encapsulation*

- **Mechanical methods:** cured epoxy does not have high mechanical strength. By applying a force, for example, using pliers, vice or awl, it is possible to bite through the epoxy material. Very often it will break off at the PCB surface or from IC package. However, in case of small SMD packages there is a high chance that some PCB components will stick to the epoxy and go off the PCB. It might be easier to control the amount of material being removed using engraving tools or CNC (computer numerical control) machine. However, there is still a danger of damaging PCB components or copper tracks.

- **Thermal methods:** cured epoxy is not very good material when it comes to elevated temperatures. Not only it starts emitting dangerous substances, but its structural strength deteriorates quickly at temperatures above 150°C (302°F). Some compounds become soft like a rubber and could be easily removed with a toothpick or needle. That way the epoxy can be easily peeled off the plastic packages of PCB components (Figure 2). However, some epoxy compounds do not soften even at 300°C (572°F), though their mechanical strength deteriorates and they become very brittle. Hence, the epoxy could be easily removed using a needle or blade. The danger of heating up the epoxy above 250°C (482°F) is associated with displacing PCB components since the solder melts at this temperature. If the epoxy is heated above 400°C (752°F) the components and PCB itself could be damaged. This is because most IC packages are made from epoxy with a filler and PCBs are normally made from FR4 which is a composite of woven fiberglass cloth with an epoxy resin binder. For precise removal of small amounts of epoxy one could use a soldering iron with a copper tip. By choosing the right tip size, applying certain force and controlling the temperature the removal process could be easily adjusted.

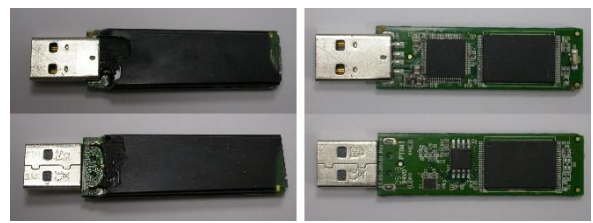

Fig. 2.  IronKey device before and after epoxy removal

- **Chemical methods:** cured epoxy is not the best material when it comes to harsh environment. It could be easily corroded by acid, alkaline, organic solvent and even water steam. If epoxy compound is immersed into specific chemical solutions the epoxy material deteriorates and softens. Then it could be scrapped off or easily removed with a blade, toothpick or needle (Figure 3). The best chemical for removing cured epoxy is nitric acid, however, it also corrodes PCB especially copper tracks and solder mask. Strong alkaline such as hot sodium hydroxide solution will corrode cured epoxy but at a much slower rate. It will also affect PCB material and IC packages. Organic solvents could work quite well and they are usually used in the industry for stripping off old paint. Some solvents could pose significant health risk and can only be used under controlled environment.

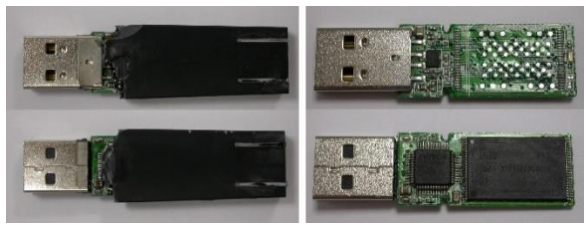

Fig. 3. IronKey device before and after epoxy removal

- **Laser ablation methods:** cured epoxy could be removed with a high power laser beam. This in fact is widely used in the industry for marking IC packages. By adjusting the energy right amount of material could be removed. However, it is quite challenging to selectively remove epoxy compound without affecting PCB components. Industrial laser cutters could still be used to quickly remove excessive epoxy compound before starting to use more precision methods.

- **Plasma methods:** cured epoxy reacts with plasma created from Ar, $O_2$ or $CF_4$ gases forming volatile compounds which are removed from the reaction chamber. This method is widely used in Failure Analysis and utilises Microwave Induced Plasma (MIP) systems [26]. By controlling the parameters different materials could be selectively removed.

By combining different methods epoxy compound can be removed very efficiently without affecting functionality of a device. For example, the excessive compound could be quickly removed using mechanical tools followed by precise removal using thermal or chemical methods. Chemical methods could also be combined with thermal ones. For example, there is an old fashioned way of removing excessive cured epoxy that involves treating it with boiling rosin also known as colophony and are still being used by plumbers to facilitate soldering of copper pipes.

### C. Identification of components

Once PCB is clear from any epoxy compound its electronic components could be identified. This enables to locate USB controller that performs on-the-fly encryption, Flash storage for firmware, applications and data, secure element responsible for hardware security protection and other components.

Passive components such as resistors (white ceramic base with black top paint), capacitors (grey or brown ceramic package) and inductors (dark brown ceramic package) could be easily spotted. It is not always easy to distinguish between capacitors and inductors, but any digital multimeter can help. Inductors have resistance close to zero and capacitors have almost infinite resistance. For more precise measurements an LCR meter must be used.

Most 2-pin, 3-pin or even 4-pin devices are likely to be diodes and transistors. Small packages will be marked with SMD (surface-mount device) codes because their full name will not fit within a small area. Their markings could be checked against SMD/SMT marking tables available from various providers. In some cases the real name of a component could be found by simple search on the Internet. Larger components are usually marked with manufacturer's logo and real name of the device under which they could be found on the manufacturer's website.

Identification becomes more challenging if the IC has custom marking. There could be two outcomes – either the manufacturer decided to disguise the real name of the device or it is actually custom made IC chip. The former could be bypassed by IC decapsulation and looking at the silicon die markings under optical microscope. In most cases the real name of the device will be clearly marked there. However, in small number of cases there could still be factory custom markings even on standard chips. In this case the device could be identified by its functionality and pinout. In worst case some possible candidates could be ordered from distributors, decapsulated and compared with the chip in question.

### D. Flash storage devices

NAND Flash chips are usually in TSOP-48 packages, but could sometime be in BGA-132, BGA-152 or LGA-52 packages. These devices have parallel interface and storage that exhibits many errors. Therefore, they usually contain additional data bits for error correction. However, that correction has to be performed by memory controller. Hence, there is always an additional cost associated with the use of NAND Flash devices. Specification of the NAND Flash is standardised by both ONFI [27] and JEDEC [28].

Secure Digital (SD) card is a proprietary non-volatile memory card format developed by the SD Association for use in portable devices [29]. SD cards exist in many different form factors, but the most popular ones are SD and microSD. The card was derived from the MultiMediaCard (MMC) and provided digital rights management based on the Secure Digital Music Initiative standard. Inside the SD card package there is a standard NAND Flash memory chip and specialised controller IC that performs error correction and Flash wear levelling [30]. Embedded MultiMediaCard (eMMC) was another development from the old MMC standardised by JEDEC [31]. It has Flash memory and controller inside a small ball grid array (BGA) IC package and partially compatible with SD cards. The eMMC interface has faster clock speed, 8-bit bus and DDR support. This allows the latest devices to achieve 3.2Gbit/s transfer rate. When it comes to security eMMC has two additional hardware security features: password protection and Replay Protected Memory Block (RPMB). The password is submitted in plaintext and not protected against eavesdropping. However, RPMB can only be read and written via successfully authenticated manner involving a random number. In addition every write access increments dedicated write counter to prevent unauthorised overwriting of the secure area.

Other Flash memory based storage devices include µSSD, UFS and NVMe (NVM Express). These are storage devices in BGA IC packages with different interfaces. Inside the package they usually have a dedicated controller chip with one or more standard NAND Flash memory chips. They all have very fast serial interfaces, however, µSSD uses SATAIII interface, UFS relies on M-PHY physical level [32] with SCSI architectural model and NVMe uses interface compatible with PCIe bus.

## IV. CHIP LEVEL TEARDOWN

This chapter introduces teardown of some encrypted USB Flash drives down to their semiconductor devices.

### A. IronKey D2, 2GB

The metal case of IronKey D2 is easy to open because it has two parts one of which is a flat lid (Figure 4). After taking off the lid the remaining body was bent with pliers and the device encapsulated in epoxy was removed from it. The epoxy compound was then removed using thermal method (Figure 5).

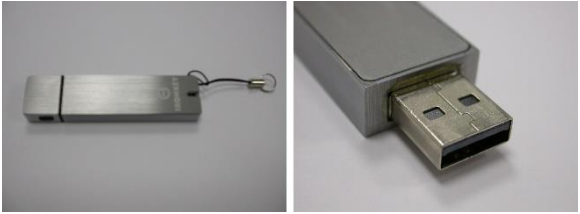

Fig. 4. IronKey D2

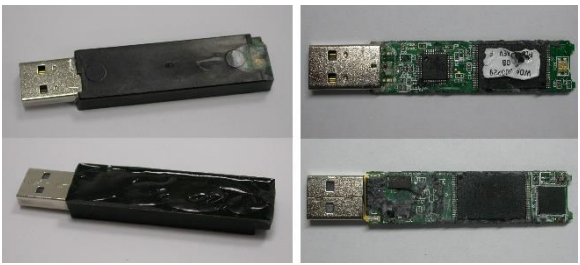

Fig. 5. IronKey D2 before and after epoxy removal

On the PCB the following main components were found: USB2.0 controller (IRONKEY 294.001), two NAND Flash chips (SAMSUNG K9K8G08U0A) and secure chip (ATMEL 98SC008CT). Figure 6 shows the die image and marking of the controller. It was designed by On Spec Inc. which in 2006 was acquired by Oban Acquisition Corp. There are only Mask ROM and SRAM areas on the die. No information is available about this chip.

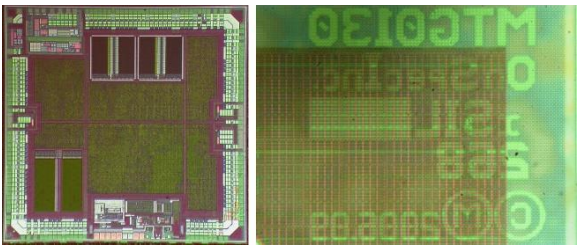

Fig. 6. Die image and marking of IRONKEY 294.001

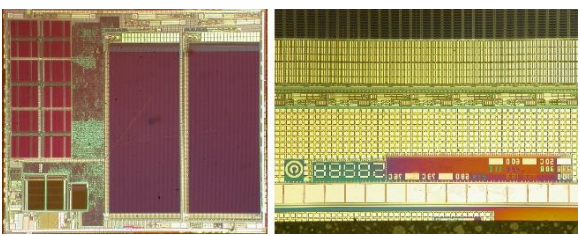

Fig. 7. Die image and marking of ATMEL 98SC008CT

Figure 7 shows the die image and marking of the secure chip. It has Mask ROM, EEPROM and SRAM on the die. This chip performs user authentication, limits the number of consecutive password attempts and transfers the data encryption key to the controller over a secure channel.

### B. IronKey D2 Version 2, 1GB

The next revision of the D2 IronKey had improved metal case. Opening it requires cutting with a mini blade (Figure 8). The epoxy was removed using chemical method (Figure 9).

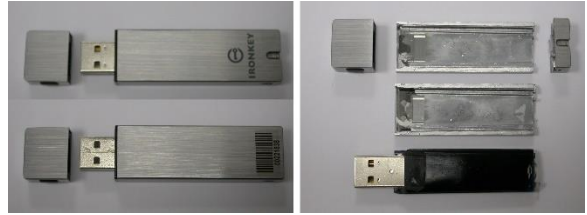

Fig. 8. IronKey D2 Ver.2 before and after opening its metal case

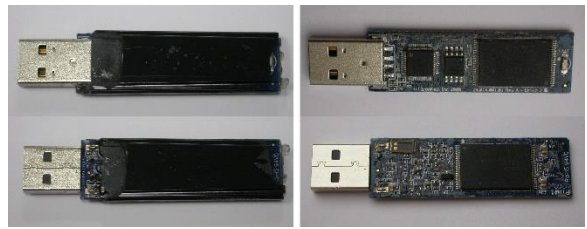

Fig. 9. IronKey D2 Ver.2 before and after epoxy removal

On the PCB the following main components were found: USB2.0 controller (IRONKEY 294.001), two NAND Flash chips (SAMSUNG K9F4G08U0A) and secure chip (ATMEL 016CU-R). It has the same controller chip as the original Ironkey D2. Figure 10 shows the die image and marking of the secure chip. It has Mask ROM, EEPROM and SRAM on the die. This chip performs user authentication, limits the number of consecutive password attempts and transfers the data encryption key to the controller over a secure channel.

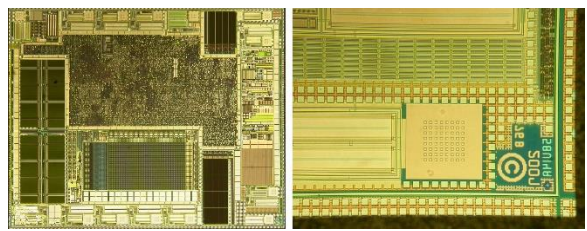

Fig. 10. Die image and marking of ATMEL 016CU-R

### C. IronKey S200, 8GB

The metal case of IronKey S200 was opened using a mini blade cutter (Figure 11). The epoxy was removed using thermal method (Figure 12).

On the PCB the following main components were found: USB2.0 controller (IRONKEY 294.005), two NAND Flash chips (SAMSUNG K9WBG08U1M) and secure chip (ATMEL 016CU-R). The controller chip was manufactured by Atmel but has only Mask ROM and SRAM areas on the die (Figure 13). It has the same secure chip as IronKey D2 Ver.2 and it performs user authentication, limits the number of consecutive password attempts and transfers the data encryption key to the controller over a secure channel.

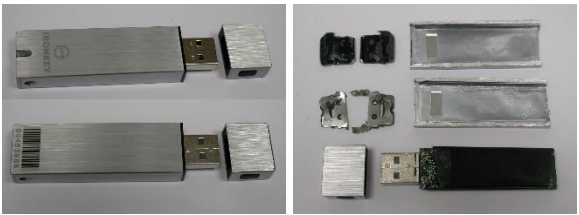

Fig. 11. IronKey S200 before and after opening its metal case

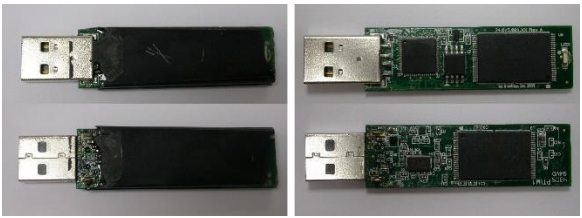

Fig. 12. IronKey S200 before and after epoxy removal

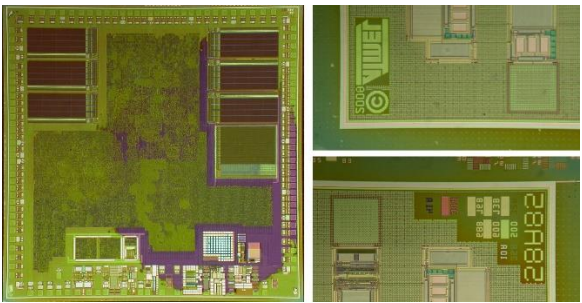

Fig. 13. Die image and marking of IRONKEY 294.005

*D. Imation IronKey F150, 2GB*

There is neither a metal case around IronKey F150 device nor its PCB is potted into an epoxy compound. Still it has FIPS 140-2 Level 3 certification. After removing metal cap at the end of the device the plastic case could be easily opened as it has two parts glued together (Figure 14).

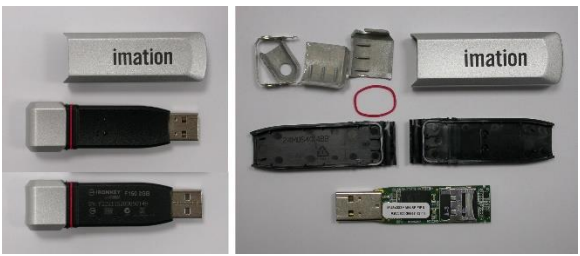

Fig. 14. IronKey F150 before and after opening its plastic case

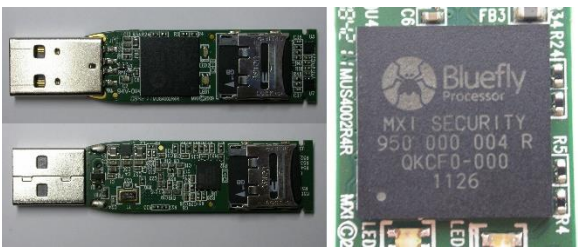

Fig. 15. PCB image and security processor of IronKey F150

Figure 15 shows the PCB of IronKey F150 and the image of its controller. This was the first IronKey device without a secure chip inside. Therefore, the encryption key and password attempts would have to be stored either in the Serial Flash chip or on the microSD card.

On the PCB the following main components were found: USB2.0 controller (Bluefly MXI SECURITY 950 000 004 R), Serial Flash chip (SST 25VF040B) and two microSD cards (1GB). The controller chip has firmware in external SPI Flash and was developed by Memory Experts International (MXI) specifically for use in encrypted USB drives and has FIPS 140-2 Level 3 certification [33].

*E. Imation IronKey D250, 4GB*

The metal case of IronKey D250 was opened using a mini blade cutter (Figure 16). The epoxy was removed using thermal method (Figure 17).

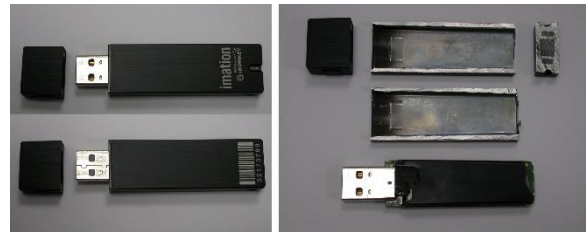

Fig. 16. IronKey D250 before and after opening its metal case

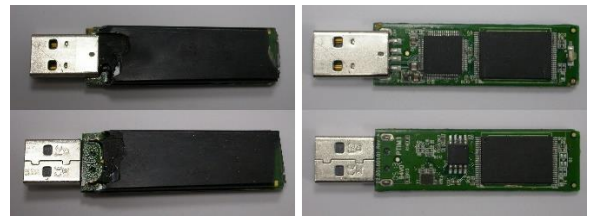

Fig. 17. IronKey D250 before and after epoxy removal

On the PCB the following main components were found: USB2.0 controller (PHISON PS2251-85-9), two NAND Flash chips (Micron 29F16G08CBACA) and secure chip (IRONKEY 31AV011). The controller chip manufactured by Phison does not have reprogrammable memory but relies on NAND Flash storage for firmware upgrades and settings. There is a proprietary utility called MPALL which sets up customised partitions, encryption settings and updates the firmware [34]. Figure 18 shows the die image and marking of the secure chip. It has Mask ROM, EEPROM and SRAM on the die. This chip performs user authentication, limits the number of consecutive password attempts and transfers the data encryption key to the controller over a secure channel.

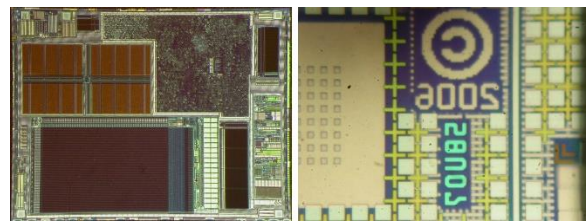

Fig. 18. Die image and marking of IRONKEY 31AV011

### F. Imation IronKey D80, 4GB

This was a budgetary version of IronKey device and another one without a metal case. However, its PCB was still potted into epoxy compound. The case was opened by cutting it with a mini saw blade (Figure 19). The epoxy was removed using chemical method (Figure 20).

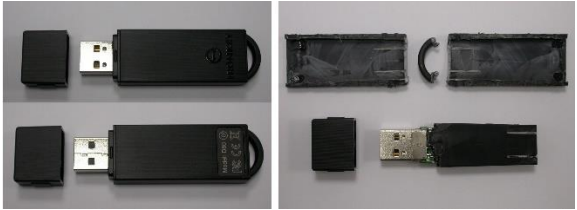

Fig. 19. IronKey D80 before and after opening its plastic case

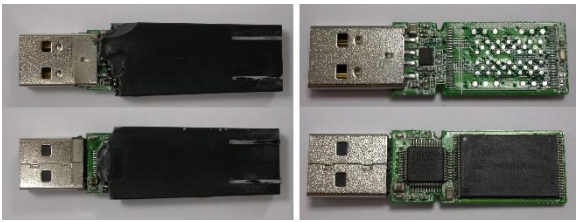

Fig. 20. IronKey D80 before and after epoxy removal

On the PCB the following main components were found: USB2.0 controller (PHISON PS2251-73-5) and NAND Flash chip (Micron 29F32G08CBACA). The controller chip manufactured by Phison relies on NAND Flash storage for firmware upgrades and settings. There is a proprietary utility called MPALL which sets up customised partitions, encryption settings and updates the firmware [34].

### G. Imation IronKey S1000, 4GB

The metal case of IronKey S1000 was opened using a mini blade cutter (Figure 21). The epoxy was removed using thermal method and the Flash storage was de-soldered during that process (Figure 22).

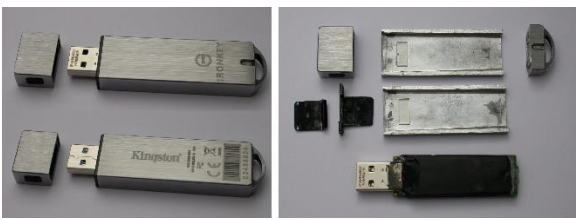

Fig. 21. IronKey S1000 before and after opening its metal case

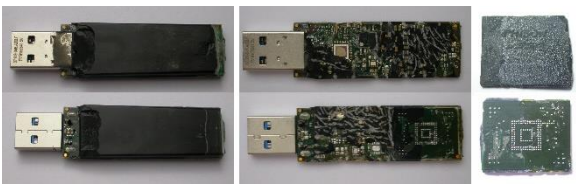

Fig. 22. IronKey S1000 before and after epoxy removal

On the PCB the following main components were found: USB3.0 SATA bridge (D720230K8), µSSD Flash chip (PHISON PSS5A311-16G) and secure chip (IRONKEY 31AV011). There is very little information about the controller chip manufactured by Renesas apart from its press release [35]. IronKey S1000 has the same secure chip as IronKey D250 and it performs user authentication, limits the number of consecutive password attempts and transfers the data encryption key to the controller over a secure channel.

### H. Kingston IronKey D300, 4GB

The metal case of IronKey D300 was opened using a mini blade cutter (Figure 23). The epoxy was removed using thermal method with mechanical force. Unfortunately, some components were physically damaged, but it was still possible to identify them by remaining IC marking (Figure 24).

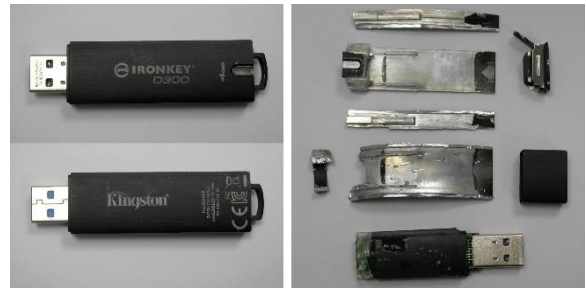

Fig. 23. IronKey D300 before and after opening its metal case

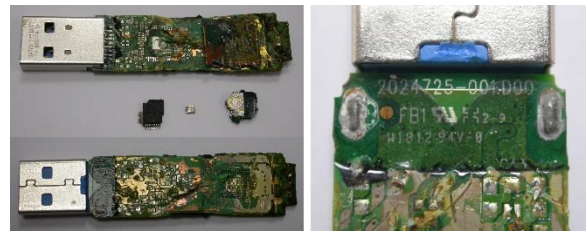

Fig. 24. IronKey D300 after epoxy removal

On the PCB the following main components were found: USB3.0 controller (PHISON PS2251-15-Q) and eMMC Flash chip (Kingston EMMC16G-TB28). The controller chip was manufactured by Phison and has only Mask ROM and SRAM areas on the die (Figure 25). Hence, it must rely on Flash storage for firmware upgrades and settings. There is a proprietary utility called MPALL which sets up customised partitions, encryption settings and updates the firmware [34].

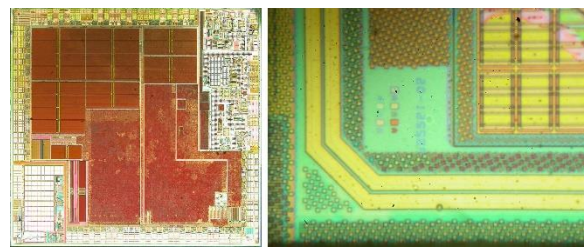

Fig. 25. Die image and marking of PHISON PS2251-15-Q

If the security features present in eMMC Flash are not used by the controller then it would be possible to reinstate the image of the storage media and bypass the limited number of consecutive incorrect password attempts [4].

### I. Kingston IronKey D300S, 4GB

The metal case of IronKey D300 was opened using a mini blade cutter (Figure 26). This device has exactly the same PCB

revision as IronKey D300. Therefore, it is expected to have the same components and level of hardware security.

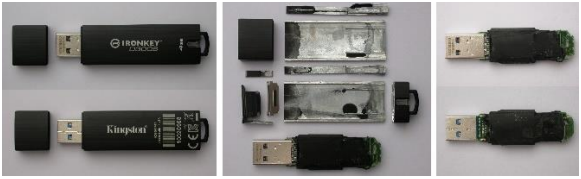

Fig. 26. IronKey D300S before and after opening its metal case

*J. Kingston DataTraveler DT4000G2, 4GB*

The plastic case of DataTraveler DT4000G2 is made from two parts clipped together and covered with metal shields. It is relatively easy to open with a knife and a flat screwdriver (Figure 27). The epoxy was removed using thermal method (Figure 28).

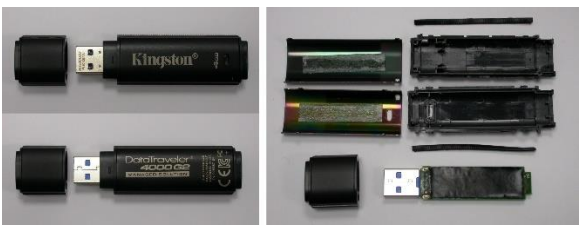

Fig. 27. DataTraveler DT4000G2 before and after opening its plastic case

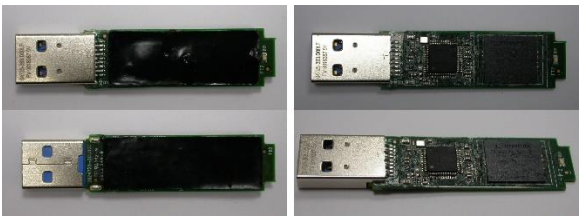

Fig. 28. DataTraveler DT4000G2 before and after epoxy removal

On the PCB the following main components were found: USB3.0 controller (PHISON PS2251-15-Q) and eMMC Flash chip (Kingston EMMC16G-TB28). These parts and PCB revision match that of IronKey D300.

*K. Kingston DataTraveler Vault Privacy 3.0, 4GB*

The plastic case of DataTraveler Vault Privacy 3.0 is made from two parts clipped together and covered with metal shields. It is relatively easy to open with a knife and a flat screwdriver (Figure 29). There is no epoxy filling inside.

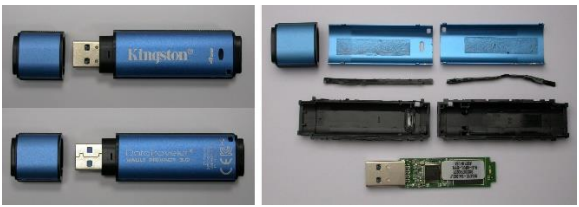

Fig. 29. DataTraveler Vault Privacy 3.0 before and after opening plastic case

On the PCB (Figure 30) the following main components were found: USB3.0 controller (PHISON PS2251-13-Q) and eMMC Flash chip (Kingston EMMC04G-M627). The PCB revision matches that of IronKey D300, but the controller chip has different marking (Figure 31). However, the die marking of the controller is exactly the same which suggests that the difference is only in firmware.

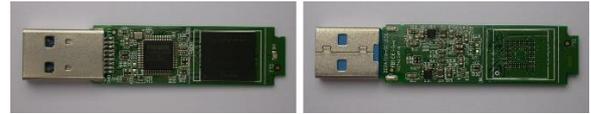

Fig. 30. PCB image of DataTraveler Vault Privacy 3.0

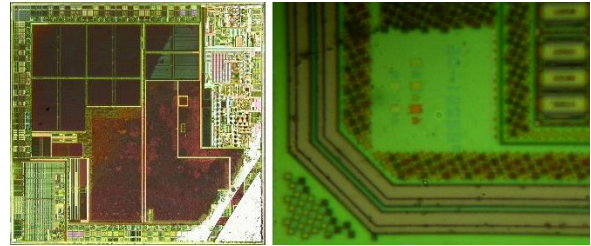

Fig. 31. Die image and marking of PHISON PS2251-13-Q

*L. Kingston DataTraveler Locker+G3, 8GB*

The metal case of DataTraveler Locker+G3 was opened using a mini blade cutter (Figure 32). The internal plastic case was easy to open with a knife. There is no epoxy filling inside.

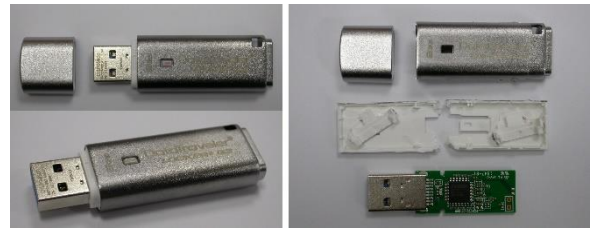

Fig. 32. DataTraveler Locker+G3 before and after opening its plastic case

On the PCB (Figure 33) the following main components were found: USB3.0 controller (PHISON PS2251-13-Q) and eMMC Flash chip (Kingston EMMC08G-M325). The controller chip is the same as in DataTraveler Vault Privacy 3.0.

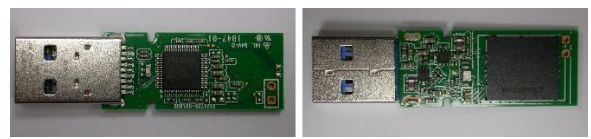

Fig. 33. PCB image of DataTraveler Locker+G3

*M. Kingston DataTraveler DT4000, 4GB*

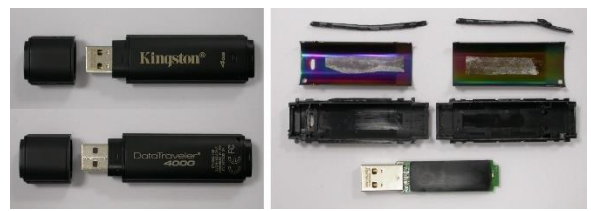

Fig. 34. DataTraveler DT4000 before and after opening its plastic case

The plastic case of DataTraveler DT4000 is made from two parts clipped together and covered with metal shields. It is relatively easy to open with a knife and a flat screwdriver

(Figure 34). The epoxy was removed using chemical method (Figure 35).

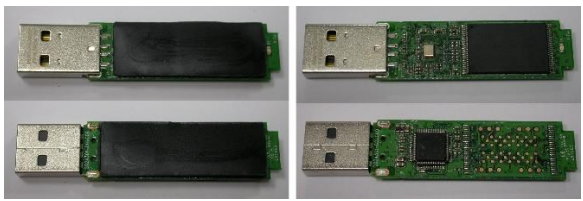

Fig. 35. DataTraveler DT4000 before and after epoxy removal

On the PCB the following main components were found: USB2.0 controller (PHISON PS2251-65-6) and NAND Flash chip (Micron 29F32G08CBACA).

*N. Kingston DataTraveler Vault Privacy, 4GB*

The plastic case of DataTraveler Vault Privacy is made from two parts clipped together and covered with metal shields. It is relatively easy to open with a knife and a flat screwdriver (Figure 36). There is no epoxy filling inside.

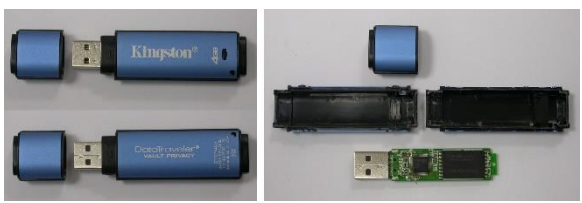

Fig. 36. DataTraveler Vault Privacy before and after opening its plastic case

On the PCB (Figure 37) the following main components were found: USB2.0 controller (PHISON PS2251-63BC-E) and NAND Flash chip (Toshiba TC58NVG4D2ETA00).

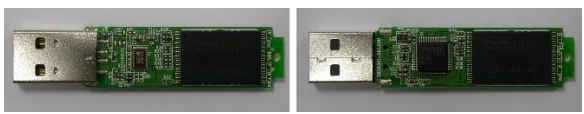

Fig. 37. PCB image of DataTraveler Vault Privacy

*O. Kingston DataTraveler DT2000, 4GB*

The metal case of DataTraveler DT2000 is easy to open because it has a plastic cap that can be mechanically removed. After that the metal case can be slid exposing the internal structure (Figure 38). After unwrapping the copper foil a keypad film was removed and a battery was de-soldered (Figure 39). Then the epoxy was removed using a combination of chemical and thermal methods (Figure 40).

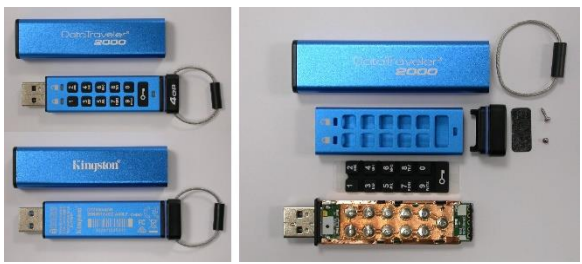

Fig. 38. DataTraveler DT2000 before and after opening its metal case

On the PCB the following main components were found: USB3.0 controller (PHISON PS2251-13-Q), eMMC Flash chip (Kingston EMMC16G-TB28) and STM32 microcontroller (L051K6). If the battery is removed after setting up the user PIN and then reinstated after a few hours the user PIN still remains active. The same applies to the number of consecutive incorrect PIN entries which by default are limited to maximum of 10.

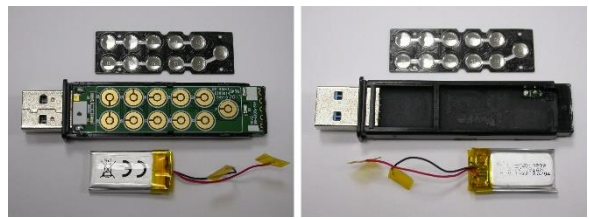

Fig. 39. DataTraveler DT2000 after disassembling the enclosure

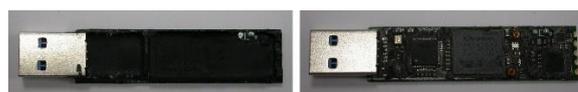

Fig. 40. DataTraveler DT2000 before and after epoxy removal

*P. iStorage Datashur Pro, 4GB*

Datashur Pro looks very similar to DataTraveler DT2000 apart from the marking. This could be explained by the fact that both Kingston and iStorage are using Datalock technology licensed from ClevX, LLC [36]. The process of disassembling Datashur Pro device is exactly the same as for DT2000 – by removing the plastic cap and sliding the metal case. Then the copper foil can be unwrapped, keypad film removed and the battery de-soldered (Figure 41). Finally the epoxy was removed using chemical method (Figure 42).

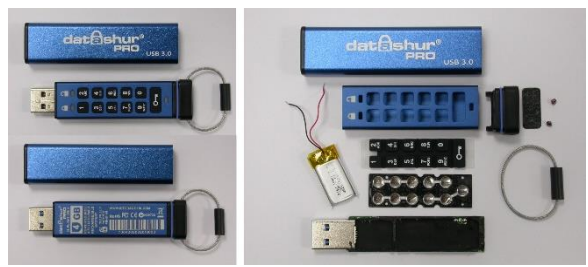

Fig. 41. Datashur Pro before and after opening its metal case

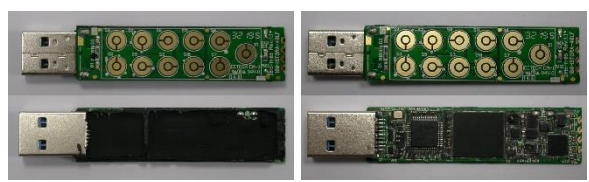

Fig. 42. Datashur Pro before and after epoxy removal

On the PCB the following main components were found: USB3.0 controller (PHISON PS2251-13-Q), eMMC Flash chip (Kingston EMMC04G-M627) and STM32 microcontroller (L051K8). Removal of the battery does not affect the user PIN and the number of consecutive incorrect PIN entries.

*Q. iStorage Datashur Pro2, 4GB*

The metal case of Datashur Pro2 was opened using a mini blade cutter (Figure 43). There are two PCBs inside – the main one with USB connector and another with keypad. The gap between PCBs is filled with epoxy compound that also covers the other side of the main PCB. The epoxy was removed using chemical method (Figure 44).

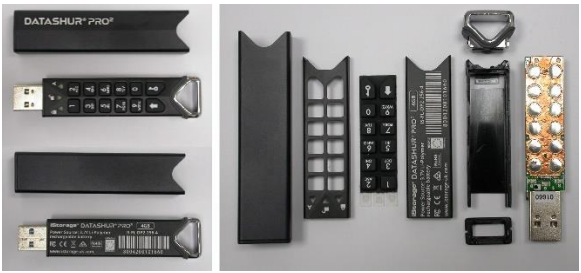

Fig. 43. Datashur Pro2 before and after opening its metal case

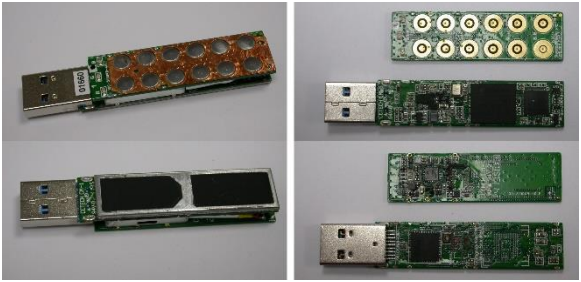

Fig. 44. Datashur Pro2 before and after epoxy removal

On the main PCB the following main components were found: USB3.1 controller (initio INIC-3861EN), eMMC Flash chip (Kingston EMMC04G-M627) and secure chip (iStorage IST61273Q). Figure 45 shows the die image and marking of the secure chip. It has Mask ROM, EEPROM and SRAM on the die. This chip performs user authentication, limits the number of consecutive password attempts and transfers the data encryption key to the controller over a secure channel. The keypad has its own controller on another PCB.

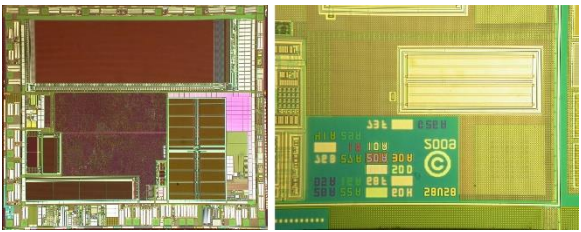

Fig. 45. Die image and marking of IST61273Q

*R. iStorage Datashur, 4GB*

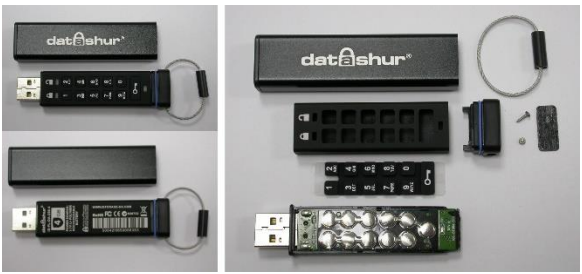

Fig. 46. Datashur before and after opening its metal case

Datashur looks very similar to Datashur Pro apart from the colour and marking. The process of its disassembling is exactly the same as for Datashur Pro – by removing the plastic cap and sliding the metal case. Then the keypad film can be removed and the battery de-soldered (Figure 46). The epoxy was removed using chemical method (Figure 47).

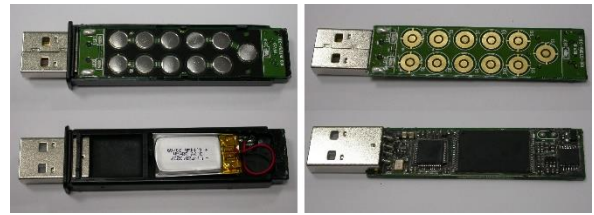

Fig. 47. Datashur before and after epoxy removal

On the PCB the following main components were found: USB2.0 controller (initio INIC-1861L), NAND Flash chip (Micron 29F32G08CBACA) and PIC16 microcontroller (L1825). Removal of the battery does not affect the user PIN and the number of consecutive incorrect PIN entries.

*S. MXI Security Stealth M200, 2GB*

The plastic case of Stealth M200 device is made from three parts glued together. It is relatively easy to open with a knife and a flat screwdriver (Figure 48). There is no epoxy filling inside. The PCB (Figure 49) looks identical to that of IronKey F150 and has the same main components: USB2.0 controller (Bluefly MXI SECURITY 950 000 004 R), Serial Flash chip (SST 25VF040B) and two microSD cards (1GB).

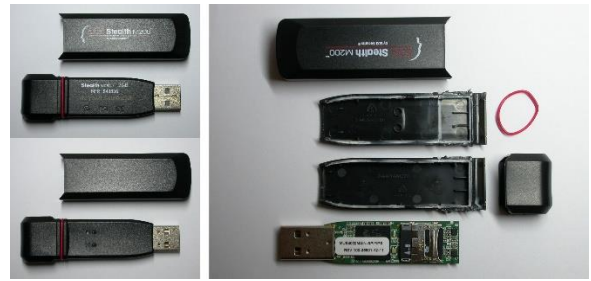

Fig. 48. Stealth M200 before and after opening its plastic case

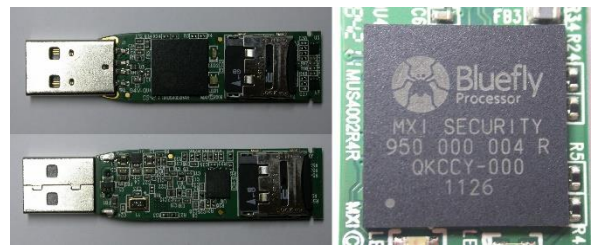

Fig. 49. PCB image and security processor of Stealth M200

*T. Integral Crypto, 8GB*

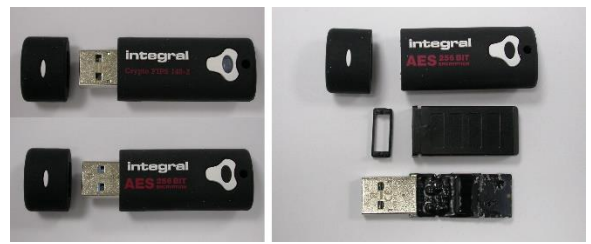

Fig. 50. Integral Crypto before and after opening its plastic case

The plastic case of Integral Crypto is protected with a rubber shield and made from two parts glued together. It is relatively easy to pull them open with a knife and a flat screwdriver (Figure 50). Then the thin layer of epoxy was removed using thermal method (Figure 51).

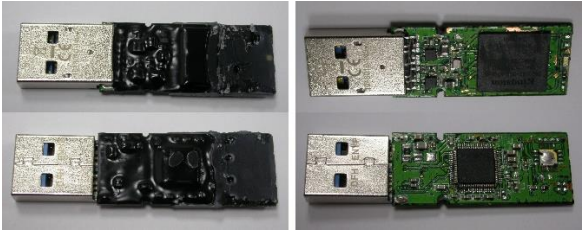

Fig. 51. Integral Crypto before and after epoxy removal

On the PCB the following main components were found: USB3.0 controller (PHISON PS2251-15-Q), eMMC Flash chip (Kingston EMMC08G-M325) and a space for secure chip in QFN-16.

*U. Integral Courier, 8GB*

The plastic case of Integral Courier is made from two parts clipped together and a small plastic piece for keyring. It is relatively easy to open with a knife and a flat screwdriver (Figure 52). There is no epoxy filling inside.

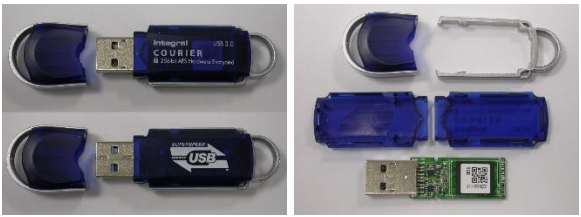

Fig. 52. Integral Courier before and after opening its plastic case

On the PCB (Figure 53) the following main components were found: USB3.0 controller (integral PS2251-13-Q), eMMC Flash chip (Kingston EMMC08G-M325) and a space for secure chip in QFN-16.

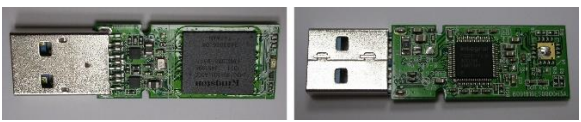

Fig. 53. PCB image of Integral Courier

*V. DataLocker Sentry 3.0, 4GB*

The metal case of DataLocker Sentry 3.0 is easy to open because it has a plastic cap that can be mechanically removed. Then the PCB can be pushed out from the USB connector side (Figure 54). There is no epoxy filling inside.

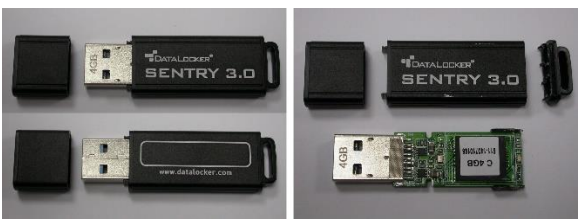

Fig. 54. Sentry 3.0 before and after opening its metal case

On the PCB (Figure 55) the following main components were found: USB3.0 controller (PHISON PS2251-13-Q), eMMC Flash chip (Kingston KE4CN2H5A) and a space for secure chip in QFN-16.

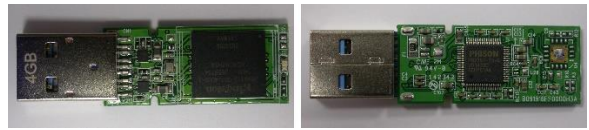

Fig. 55. PCB image of Sentry 3.0

*W. DataLocker Sentry, 4GB*

The metal case of DataLocker Sentry is easy to open because it has a plastic cap that can be mechanically removed. Then the PCB can be pushed out from the USB connector side (Figure 56). There is a fake epoxy filling inside which in fact is a black paint. It was removed with a DIY paint cleaner (Figure 57).

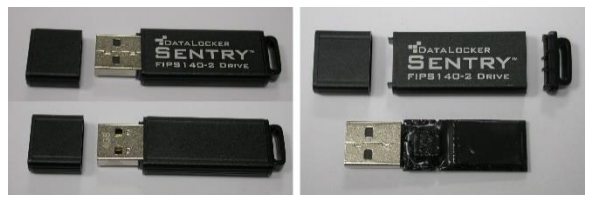

Fig. 56. Sentry before and after opening its metal case

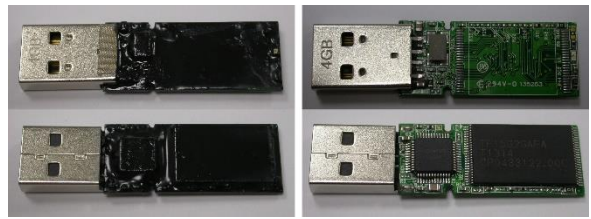

Fig. 57. Sentry before and after fake epoxy removal

On the PCB the following main components were found: USB2.0 controller (BLOCKMASTER BM9931) and NAND Flash chip (TF15G2GAFA). The controller chip was manufactured by BlockMaster and has only Mask ROM and SRAM areas on the die (Figure 58). Hence, it must rely on Flash storage for firmware upgrades and settings. There is no public information about this USB controller.

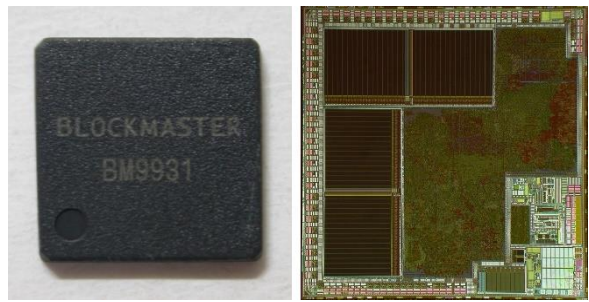

Fig. 58. Package marking and die image of BLOCKMASTER BM9931

*X. SafeXS Protector, 4GB*

The plastic case of SafeXS Protector is made from two parts glued together and covered with a metal shield. It is relatively easy to open with a knife and a flat screwdriver (Figure 59). There is no epoxy filling inside.

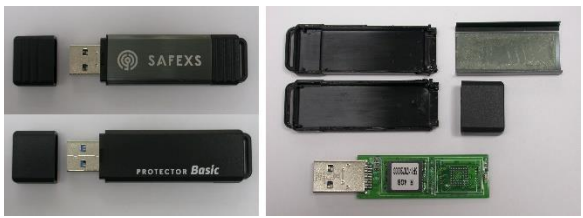

Fig. 59. SafeXS Protector before and after opening its plastic case

On the PCB (Figure 60) the following main components were found: USB3.0 controller (PHISON PS2251-13-Q), eMMC Flash chip (Kingston EMMC04G-M627) and a space for secure chip in QFN-16.

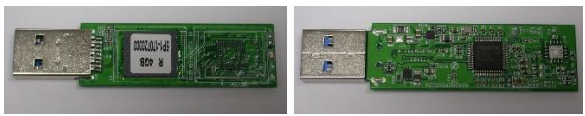

Fig. 60. PCB image of SafeXS Protector

V. ATTACK METHODS

The only attacks carried out during the research described in this paper were related to disabling the limited number of password retry attempts. Three different devices with microSD, NAND Flash and eMMC Flash were tested against NAND mirroring attacks [4]. In addition the security of eMMC chips was evaluated as well as possible attack vectors on Atmel AT90SC family of chips.

*A. Imation IronKey F150*

This device has two types of Flash storage on its PCB: Serial Flash chip and microSD cards.

The Serial Flash chip was de-soldered from the PCB and placed into a ZIF socket (Figure 61, left). This allowed monitoring of its contents using universal programmer. However, no changes were observed inside this chip during the device operation with different number of incorrect consecutive password attempts.

The microSD card adapter was made for the use with EasyJTAG plus tool [37]. This is a very powerful tool when it comes to working with raw memory images. Then the images of both 1GB microSD cards used in IronKey F150 were made with different number of failed password attempts (Figure 61, right).

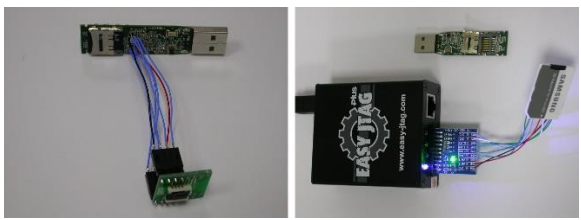

Fig. 61. Testing Serial Flash and microSD card from IronKey F150

By restoring the memory image from the rear side microSD card back to its original state it was possible to reinstate the remaining number of password attempts back to ten. This proved that IronKey F150 is vulnerable to NAND mirroring attacks despite to its FIPS 140-2 Level 3 certification.

*B. Imation IronKey D80*

This device has only one type of Flash storage on its PCB – Parallel NAND Flash chip.

The image of the 4GB Micron NAND Flash chip was created using the same EasyJTAG plus tool with a NAND TSOP-48 adapter (Figure 62).

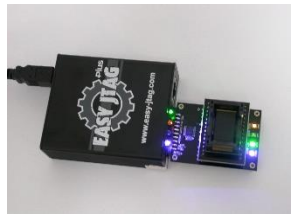

Fig. 62. Testing NAND Flash from IronKey D80

By restoring the memory image back to its original state and soldering it back to the PCB it was possible to reinstate the remaining number of password attempts back to ten. This proved that IronKey D80 is vulnerable to NAND mirroring attacks.

*C. Kingston DataTraveler DT4000G2 identical to IronKey D300 and IronKey D300S*

These devices have only one type of Flash storage on their PCB – eMMC flash chip. Revisions of their PCBs are exactly the same and, therefore, their hardware security protection is assumed to be the same. The actual NAND mirroring attempt was performed on Kingston DataTraveler DT4000G2 device.

The PCB was cleared from epoxy compound using thermal method. Then the eMMC chip was desoldered and cleaned from remaining solder (Figure 63).

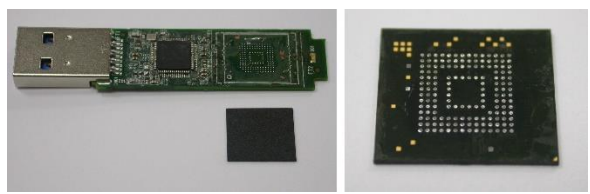

Fig. 63. Testing eMMC Flash from DataTraveler DT4000G2

The image of the 16GB Kingston eMMC Flash chip was created using the same EasyJTAG plus tool with an eMMC-153 adapter. However, on practice only the first 4GB of its ROM1 partition was used for data storage. Neither its password protection feature, nor RPMB partition were activated. After that the eMMC chip was re-balled using a stencil, 0.25mm solder balls and hot air gun before being soldered back on cleaned DT4000G2 PCB (Figure 64).

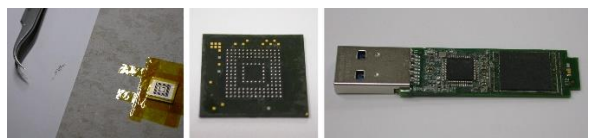

Fig. 64. Soldering eMMC Flash back to PCB of DataTraveler DT4000G2

The above process was repeated after exhausting 8 out of 10 incorrect consecutive password attempts. Then the memory image inside the eMMC chip was restored back to its original state and the chip was soldered back to the PCB. That way it was possible to reinstate the remaining number of password attempts back to ten. This proved that DataTraveler DT4000G2 is vulnerable to NAND mirroring attacks. Very likely IronKey D300 and D300S will be vulnerable to NAND mirroring attacks as well, because they share the same PCB revision and components.

*D. eMMC Flash devices*

eMMC devices comply to a dedicated JEDEC standard that covers their pinout, communication protocol and security features [31]. eMMC Flash storage devices have specialised controller and one or more standard NAND Flash chips inside BGA package. The controller not only performs error correction and NAND Flash wear levelling, but offers some security features. Those involve possibility to protect access to the device with user password and specialised RPMB partition with symmetric authentication key and proprietary algorithm. Upon activation with one-time programming (OTP) key any access to this partition will require authentication with challenge-response protocol involving random number to prevent replay attacks. Any write access to this partition increases a counter thus making it impossible to modify its contents without being detected.

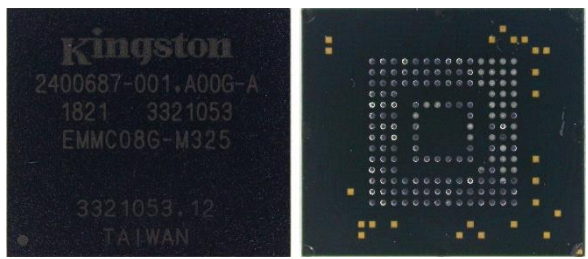

Fig. 65. Kingston EMMC08G-M325 from both sides

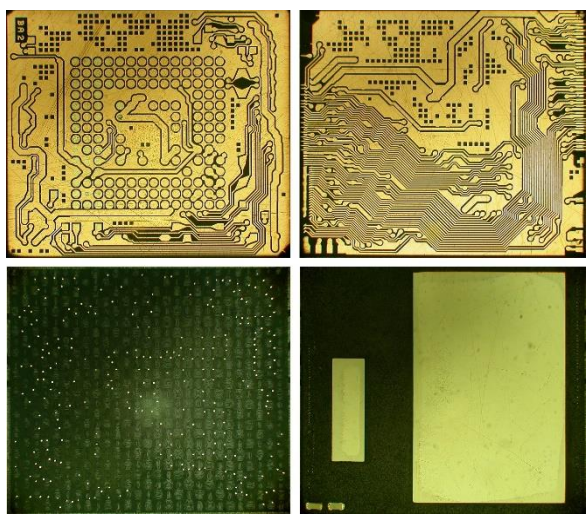

Fig. 66. EMMC08G-M325 polished to solder, vias, components and dies

One of eMMC chips used in Kingston DataTraveler Locker+G3 USB drive, EMMC08G-M325, was investigated for possible ways of bypassing hardware security protection features. Figure 65 shows picture of the eMMC chip from both sides. Apart from the standard 153 balls present on the chip there are 28 square pads. In order to investigate their connections the bottom side of the chip was slowly polished down to the silicon dies (Figure 66). It can be observed that all the traces on the BGA carrier PCB between the controller (small die) and the NAND Flash (large die) have connections with the exposed pads. By tapping those pads with a logic analyser it would be possible to eavesdrop on the communication between the controller and the NAND memory chip. However, we need to be sure that no security information is stored inside the controller chip and that the NAND die is a standard NAND memory.

By polishing the eMMC chip further and chemically deprocessing the remaining bulk silicon the internal structures of the controller chip and the NAND chip were exposed (Figure 67). Their careful investigation under a microscope revealed that there is no non-volatile reprogrammable memory such as EEPROM or Flash present in the controller. It has a Mask ROM which is used to store the firmware or it could load the firmware from a specific area in the NAND chip.

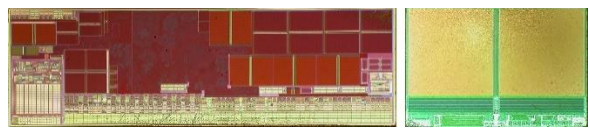

Fig. 67. Controller and NAND Flash dies inside EMMC08G-M325

The investigation presented above shows how the security features of eMMC chip could be reinstated back to their original state by restoring the content of the NAND Flash chip inside eMMC device. The eMMC controller has a reset input, hence, it would be relatively easy to disable it during the direct communication with the internal NAND chip. Since most NAND Flash devices are compliant to ONFI standard [27] it should not take much effort to figure out the manufacturer, device type and other features.

*E. Atmel secure microcontrollers*

In Chapter 4 some Atmel devices were found to be used in many IronKey products and one iStorage product. Now more attention will be paid towards their proper identification.

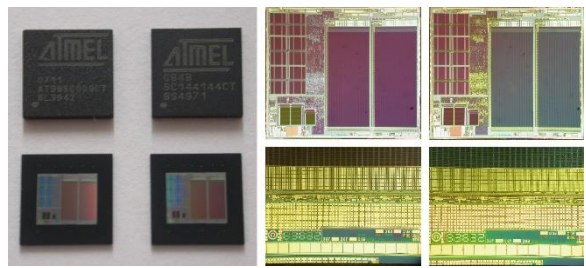

Fig. 68. Die images and marking of AT98SC008CT and AT90SC144144CT

The secure chip used in IronKey D2 was marked as 98SC008CT and this suggests that its full name is AT98SC008CT [38]. At the same time the die has the following marking: 58888. By searching on the Internet it was found that the similar marking has another Atmel device AT90SC144144CT [39] which is a reconfiguration of AT90SC320288CT device [40]. In order to investigate this further blank samples of AT98SC008CT and AT90SC144144CT devices were ordered. The die images of these chip and die markings are exactly the same (Figure 68) and also match that of the secure chip inside IronKey D2. This suggests that AT98SC008CT chip is actually based on AT90SC144144CT or AT90SC320288CT chip with modified firmware.

The secure chips used in IronKey D2 Ver.2 and IronKey S200 were marked as 016CU-R. This suggests that the full name of this chip is AT98SC016CU [41]. At the same time the die has the following marking: 58U14A. By searching on the Internet it was found that the similar marking has another Atmel device AT90SC12818RCU [42]. In order to investigate this further blank samples of AT98SC016CU devices were ordered. The die image of this chip and die marking (Figure 69) fully match those of the secure chips used in IronKey D2 Ver.2 and S200 devices. This suggests that AT98SC016CU chip is most likely based on AT90SC12818RCU chip with modified firmware.

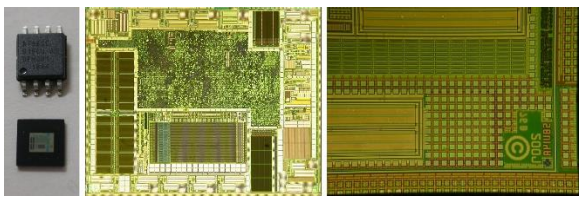

Fig. 69. Die image and marking of AT98SC016CU

The secure chip used in IronKey D250 and IronKey S1000 was marked as IRONKEY 31AV011. However, its die has the following marking: 58U07. By searching on the Internet it was found that the similar marking has another Atmel device AT90SC28872RCU [43] also marketed as AT90SC28848RCU [44]. Very likely that this chip was programmed by IronKey to perform user authentication, limit the number of password attempts and encryption key secure storage.

The secure chip used in iStorage Datashur Pro2 was marked as iStorage IST61273Q. However, its die has the following marking: 58U58. By searching on the Internet it was found that the similar marking has another Atmel device AT90SO128 [45]. Very likely that this chip was programmed by iStorage to perform user authentication, limit the number of password attempts and encryption key secure storage.

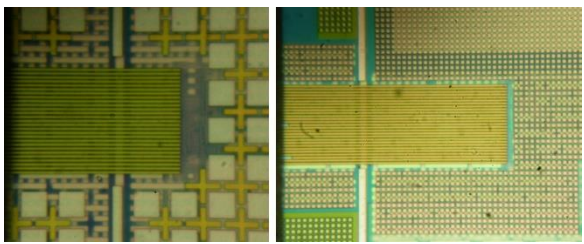

Fig. 70. Silicon level backdoor present in many AT90SC devices

One possibility for attacking AT90SC family of devices could be through the silicon level backdoor or factory debug interface. All AT90SC secure chip devices found in IronKey and iStorage products have recognisable feature on their die in the form of 27 wires going outside the perimeter of the device (Figure 70). If this interface is understood, for example through partial reverse engineering of the on-chip logic, it could be exploited for full access to embedded memory and CPU registers. However, for establishing access to those wires some test points created by using FIB (Focused Ion Beam) machine [46] will be required. The challenge associated with any chip editing of AT90SC devices could come from the presence of top layer sensor mesh (Figure 71). It has to be either disabled or bypassed with rewiring.

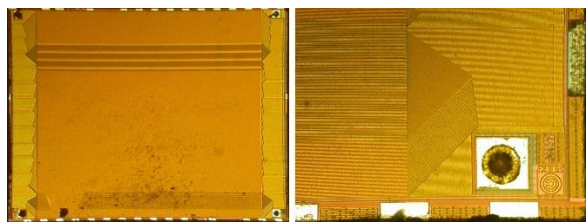

Fig. 71. Die image of IRONKEY 31AV011 and sensor mesh near marking

Another possibility of getting access to the firmware and data is through possible backdoors in firmware or in user applications. To achieve this the firmware stored in Mask ROM should be extracted first, however, it is usually trivial to extract the content of Mask ROM. For particular AT90SC devices fabricated with 180nm and 150nm process the bits in Mask ROM should be visible under optical microscope. For example, this is the case for secure chip used in IronKey S1000 (Figure 72). However, some descrambling of the data will have to be performed before the firmware could be disassembled and analysed.

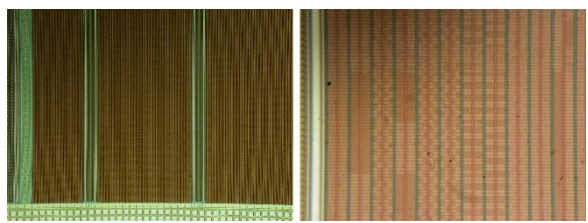

Fig. 72. Mask ROM data visible under optical microscope in AT90SC chip

Data extraction from EEPROM to access user applications and possible key storage is a bit trickier. The example of the advanced contrast SEM imaging applied to the AT98SC008CT chip is presented in Figure 73. The difference between erased memory cell (in blue circle) and programmed cell (red circle) can be clearly determined. This method proved to be working well for custom designed chips such as in medical devices [47]. The data bus of the array is 16-bit wide and there is no error correction. Hence, for reliable extraction the success rate must be above 99.99%.

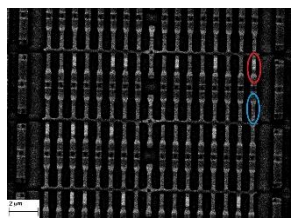

Fig. 73. Advanced contrast image of EEPROM in AT98SC008CT

Other attack methods on AT90SC devices could involve laser fault injection attacks [48] and electromagnetic fault injection [49].

More advanced methods of attacking many secure chips involve reverse engineering of the silicon logic, data extraction from Mask ROM and EEPROM and subsequent simulation of the device using EDA (Electronic Design Automation) tools.

## VI. COMPARISON OF DEVICES

This chapter summarises the findings in encrypted USB Flash drives and compares their hardware security level of protection. Table I outlines the findings described in Sections 4 and 5 on encrypted USB Flash drives from IronKey, Imation, Kingston, iStorage, MXI, DataLocker, Integral and SafeXS.

TABLE I. COMPARISON OF ENCRYPTED USB FLASH DRIVES

| Name | Features | | | | |
|---|---|---|---|---|---|
| | PCB revision | Cryptochip (AES256) | Secure Element | Flash | Certification |
| IronKey D2 USB2.0 | Drive V1 IronKey 2006 Epoxy compound | IRONKEY 294.001 (On Spec Inc) | AT98SC008 | Samsung NAND | FIPS140-2 Level 2 EAL5+ |
| IronKey D2 Ver.2 USB2.0 | 24.014.001.01 Drive 2 IronKey 2008 Epoxy compound | IRONKEY 294.001 (On Spec Inc) | AT98SC016 | Samsung NAND | FIPS140-2 Level 2 EAL4+ |
| IronKey S200 USB2.0 | 24.015.001.XX IronKey 2009 Epoxy compound | IRONKEY 294.005 (Atmel) | AT98SC016 | Samsung NAND | FIPS140-2 Level 3 EAL4+ |
| IronKey F150 USB2.0 | MUS4002R4R MXI 2009 | Bluefly MXI SECURITY 950 000 004 R | N/A | Samsung microSD | FIPS140-2 Level 3 |
| IronKey D250 USB2.0 | 22.051.001.01 Rev A 0513 Epoxy compound | PHISON PS2251-85-9 | IRONKEY 31AV011 (Atmel) | Micron NAND | FIPS140-2 Level 3 EAL5+ |
| IronKey D80 USB2.0 | P091538FP242401A | PHISON PS2251-73-5 | N/A | Micron NAND | N/A |
| IronKey S1000 USB3.1 | 2018 Epoxy compound | Renesas µPD720230 | IRONKEY 31AV011 (Atmel) | Phison µSSD SATAIII | FIPS140-2 Level 3 EAL5+ |
| IronKey D300 USB3.1 | 2024725-001.D001 1812 Epoxy compound | PHISON PS2251-15-Q | N/A | Kingston eMMC | FIPS140-2 Level 3 |
| IronKey D300S USB3.1 | 2024725-001.D00 2008 Epoxy compound | PHISON PS2251-15-Q | N/A | Kingston eMMC | FIPS140-2 Level 3 |
| Kingston DT4000G2 USB3.1 | 2024725-001.D00 2013 Epoxy compound | PHISON PS2251-15-Q | N/A | Kingston eMMC | FIPS140-2 Level 3 |
| Kingston DT Vault Privacy 3.0 USB3.1 | 2024725-001.D00 1742 | PHISON PS2251-13-Q | N/A | Kingston eMMC | FIPS197 |
| Kingston DT Locker+G3 USB3.1 | 2024726-001.B00 1847 | PHISON PS2251-13-Q | N/A | Kingston eMMC | N/A |
| Kingston DT4000 USB2.0 | 2024301-001.C00 1501 Epoxy compound | PHISON PS2251-65-6 | N/A | Micron NAND | FIPS140-2 Level 2 |
| Kingston DT Vault Privacy USB2.0 | 2024301-001.B00 | PHISON PS2251-63BC-E | N/A | Toshiba NAND | N/A |
| Kingston DT2000 USB3.1 | 100-ISTORA-61LF Epoxy compound | PHISON PS2251-13-Q | N/A | Kingston eMMC | FIPS140-2 Level 3 |
| iStorage Datashur Pro USB3.2 | 100-ISTORA-40LF 3219 Epoxy compound | PHISON PS2251-13-Q | N/A | Kingston eMMC | FIPS140-2 Level 3 |
| iStorage Datashur Pro2 USB3.2 | 100-3100MB-60LF 2419 Epoxy compound | Initio INIC-3861EN | iStorage IST61273Q (Atmel) | Kingston eMMC | FIPS140-2 Level 3 EAL5+ |
| iStorage Datashur USB2.0 | 100-ISTORA-31LF Epoxy compound | Initio INIC-1861L | N/A | Micron NAND | FIPS140-2 Level 3 |
| MXI Stealth M200 USB2.0 | MUS4002R4R MXI 2009 | Bluefly MXI SECURITY 950 000 004 R | N/A | Samsung microSD | FIPS140-2 Level 3 |
| Integral Crypto USB3.0 | B091916ES0000H3A Epoxy compound | PHISON PS2251-15-Q | N/A | Kingston eMMC | FIPS140-2 Level 2 |
| Integral Courier USB3.0 | B091916ES0000H3A | PHISON PS2251-13-Q | N/A | Kingston eMMC | FIPS197 |
| DataLocker Sentry 3.0 USB3.0 | B091916ES0000H3A | PHISON PS2251-13-Q | N/A | Kingston eMMC | FIPS197 |
| DataLocker Sentry USB2.0 | 135263 Fake epoxy (paint) | BlockMaster BM9931 | N/A | TF15G2GAFA | FIPS140-2 Level 2 |
| SafeXS Protector USB3.0 | B018216EP0000G5A | PHISON PS2251-13-Q | N/A | Kingston eMMC | N/A |

As it could be observed from the comparison table, very limited number of USB drives use a secure chip for protecting passwords and encryption key. Even some drives with a high FIPS 140-2 Level 3 certification do not have secure chip inside. This could potentially open all sorts of attacks on inadequately protected devices.

## VII. FURTHER WORK

So far only devices without secure elements were demonstrated to be vulnerable to NAND mirroring attacks [4]. This work could be extended with full off-line password brute forcing by understanding how the encryption key storage is encrypted with the user password.

The next step could be in mounting attacks against Atmel secure microcontrollers used in IronKey and iStorage devices.

## VIII. CONCLUSION

This paper describes the teardown process of nine generations of IronKey encrypted USB Flash drives from the very first model to the latest one. Out of them four models were found vulnerable to NAND mirroring attacks. This includes the latest model certified to FIPS 140-2 Level 3 and approved by NATO.

The changes between devices in IronKey family were discussed and it was revealed that some members do not have adequate protection against possible attacks. Even the devices with secure microcontrollers should be possible to attack with modern attack methods.

Since the IronKey technology changed its hands over time, devices from associated manufacturers were analysed as well. In addition, similar devices from some other manufacturers were analysed to give a broad picture of the situation in the area of encrypted USB Flash storage devices. This resulted in the teardown of 15 different models of encrypted USB Flash drives from Kingston, iStorage, MXI, DataLocker, Integral and SafeXS. Most of these devices were found to be potentially vulnerable to attacks.

Surprisingly none of the security features available in eMMC chips were used in IronKey and other devices. Neither the advantage of the battery in Kingston DT2000 and iStorage devices was used to improve their security by storing the keys in battery-backed SRAM. The research described in this paper demonstrated that some secure devices marketed as unique are in fact software implementation of functions on a different secure microcontroller of standard family.

The users of encrypted USB Flash drives must be aware of the hardware security features implemented in devices they use. This is especially important for sensitive data in military, government and corporate applications. This paper sheds some light on the actual situation with security of encrypted USB Flash drives. However, it only does this to the level of feasibility study against possible attacks. For example, it might not explain how some secret data could have leaked from encrypted USB Flash drives. If the hardware security of IronKey devices is weak this could possibly be exploited by attackers. More research will be needed to make sure that the hardware security of semiconductor devices used in IronKey USB Flash drives meet the highest expectations.

## ACKNOWLEDGMENT

Author would like to thank Dr Olga Larina for inspiring fruitful discussions and helpful assistance.

# REFERENCES

[1] IronKey, Crypto Museum. www.cryptomuseum.com, February 2018
[2] S. Skorobogatov, "Flash Memory 'Bumping' Attacks," Cryptographic Hardware and Embedded Systems Workshop (CHES), August 2010, LNCS 6225, Springer, ISBN 3-642-15030-6, pp.158-172
[3] S. Skorobogatov, and C. Woods, "Breakthrough silicon scanning discovers backdoor in military chip," Cryptographic Hardware and Embedded Systems Workshop (CHES), September 2012, Leuven, Belgium, LNCS 7428, Springer, ISBN 978-3-642-33026-1, pp.23-40
[4] S. Skorobogatov, "The bumpy road towards iPhone 5c NAND mirroring," arXiv:1609.04327, September 2016
[5] F. Courbon, S. Skorobogatov, and C. Woods, "Reverse engineering Flash EEPROM memories using Scanning Electron Microscopy," In Proceedings of the 15th Smart Card Research and Advanced Application Conference (CARDIS), Cannes, France, November 2016
[6] S. Skorobogatov, "Is Hardware Security prepared for unexpected discoveries?" 25th International Symposium on the Physical and Failure Analysis of Integrated Circuits (IPFA), July 2018, Singapore. IEEE Xplore 2018
[7] https://www.ironkey.com/en-US/solutions/golden-ironkey.html
[8] S. Skorobogatov, "Compromising device security via NVM controller vulnerability," IEEE International Conference on Physical Assurance and Inspection of Electronics (PAINE), December 2020. IEEE Xplore
[9] D. Jagos, "Security analysis of USB drive," Master's Thesis, Faculty of Information Technology, CTU in Prague, February 2018
[10] J.-M. Picod, R. Audebert, S. Blumenstein, and E. Bursztein, "Attacking encrypted USB keys the hard(ware) way," Research at Google
[11] M. Sparrow, "Keep Your Data From Prying Eyes With This Military-Grade IronKey USB Drive," Forbes, May 2020
[12] N. Popper, "Lost Passwords Lock Millionaires Out of Their Bitcoin Fortunes," The New York Times, January 2021
[13] Compliance FAQs: Federal Information Processing Standards (FIPS). National Institute of Standards and Technology (NIST), July 2021
[14] Common Criteria for Information Technology Security Evaluation. Common Criteria, Revision 5, April 2017
[15] M.J. Dworkin, E.B. Barker, J.R. Nechvatal, J. Foti, L.E. Bassham, E. Roback, and J.F. Dray Jr., "Advanced Encryption Standard (AES)", FIPS PUB 197, NIST, November 2001
[16] A. Lee, M.E. Smid, and S.R. Snouffer, "Security Requirements for Cryptographic Modules," FIPS PUB 140-2, NIST, May 2001
[17] Security Requirements for Cryptographic Modules. FIPS PUB 140-3, NIST, March 2019
[18] L. Mearian, "Imation buys IronKey's hardware assets," News, Computerworld, September 2011
[19] Imation Acquires MXI Security. Businesswire, June 2011
[20] Imation Adds IronKey S1000 Products to World's Most Secure Family of USB Storage Devices – Now Also the World's Fastest Hardware-Encrypted USB Flash Drives. Businesswire, February 2015
[21] DataLocker Acquires IronKey Enterprise Management Services (EMS) and Other Assets From Imation. News & Events, February 2016
[22] A. Armstrong, "Kingston Acquires IronKey USB Technology From Imation," StorageReview, February 2016
[23] Kingston Technology Company, Inc. DataTraveler DT4000 G2 Series USB Flash Drive. FIPS 140-2 Non-Proprietary Security Policy, December 2014
[24] Kingston Technology Company, Inc IronKey D300 Series USB Flash Drive. FIPS 140-2 Non-Proprietary Security Policy, February 2018
[25] Kingston Introduces New IronKey D300S Encrypted USB. Tech Critter, News, November 2018
[26] J. Tang, B. Wang, C. Liu, J. Wang, and C. I. M. Beenakker, "Unique Failure Analysis Capabilities Enabled by the MIP Decapsulation Technique," in Proceedings 24th International Symposium on the Physical and Failure Analysis of Integrated Circuits (IPFA), 2017
[27] Open NAND Flash Interface. ONFI Specifications http://www.onfi.org/specifications
[28] JESD230C, NAND Flash Interface Interoperability. November 2016
[29] SD Standards – Universal High Performance Mobile Storage. SD Association, July 2021
[30] Are you using the right SD card? Simms International plc, DesignSpark, November 2017
[31] JESD84-B51A: Embedded MultiMediaCard (e·MMC). Electrical Standard (5.1A), January 2019
[32] MIPI M-PHY v5.0. MIPI Alliance, September 2021
[33] Bluefly Processor, Security Policy. Memory Experts International, April 2010
[34] Phison MPALL v5.13.0C. USBDEV. https://www.usbdev.ru/files/phison/mpall/
[35] Renesas Electronics Introduces World's First USB 3.0-SATA3 Bridge SoC Supporting High-Speed UASP Protocol. Renesas, August 2011
[36] Kingston releases encrypted USB with keypad access. Help Net Security, January 2016
[37] EasyJTAG plus. https://easy-jtag.com/
[38] Secure ASSP AT98SC008CT Summary. Atmel, January 2006
[39] Secure Microcontroller for Smart Cards AT90SC144144CT Summary. Atmel, September 2005
[40] Secure Microcontroller for Smart Cards AT90SC320288RCT Summary. Atmel, April 2005
[41] Secure ASSP AT98SC016CU Summary. Atmel, June 2007
[42] Security Target Lite AT90SC12818RCU. Common Criteria
[43] Secure Microcontroller for Smart Cards AT90SC28872RCU Summary. Atmel, May 2007
[44] Secure Microcontroller for Smart Cards AT90SC28848RCU Summary. Atmel, August 2008
[45] Secure Microcontroller for Security Modules AT90SO128 Preliminary Summary. Atmel, January 2010
[46] O. Kömmerling, and M.G. Kuhn, "Design principles for tamper-resistant smartcard processors," USENIX Workshop on Smartcard Technology, Chicago, Illinois, USA, May 1999
[47] S. Skorobogatov, "Deep dip teardown of tubeless insulin pump," arXiv:1709.06026, September 2017
[48] S. Skorobogatov, and R. Anderson, "Optical Fault Induction Attacks," Cryptographic Hardware and Embedded Systems Workshop (CHES), August 2002, LNCS 2523, Springer-Verlag, pp.2-12
[49] S. Skorobogatov, "Hardware Security Evaluation of MAX 10 FPGA: Feasibility Study of Intel® MAX 10 devices for compliance to MODH security level," arXiv:1910.05086, October 2019